\documentclass[10pt,a4paper,twocolumn]{elsart5p}

\usepackage{graphicx}
\usepackage{latexsym}   
\usepackage{lineno}

\begin{document}


\journal{Physics Letters A}

\begin{frontmatter}

\title{Arithmetical and geometrical means of generalized logarithmic and exponential functions: generalized sum and product operators}

\author[Filo]{Tiago Jos\'e Arruda}, 
\author[Filo]{Rodrigo Silva Gonz\'alez},  
\author[Filo,Barao]{C\'esar Augusto Sangaletti Ter\c{c}ariol}, 
\author[Filo]{Alexandre Souto Martinez\corauthref{cor1}}
\corauth[cor1]{\ead{asmartinez@ffclrp.usp.br}}
\address[Filo]{Faculdade de Filosofia, Ci\^encias e Letras de Ribeir\~ao Preto (FFCLRP) \\
         Universidade de S\~ao Paulo (USP) \\
         Av.~Bandeirantes, 3900 \\
         14040-901  Ribeir\~ao Preto, SP, Brazil}
\address[Barao]{Centro Universit\'ario Bar\~ao de Mau\'a \\
         Rua Ramos de Azevedo, 423 \\ 
         14090-180, Ribeir\~ao Preto, SP, Brazil}
\date{\today}


\begin{abstract}
One-parameter generalizations of the logarithmic and exponential functions have been obtained as well as algebraic operators to retrieve extensivity. Analytical expressions for the successive applications of the sum or product operators on several values of a variable are obtained here. Applications of the above formalism are considered. 
\end{abstract}

\end{frontmatter}

\section{Introduction}

Based on non-extensive thermostatistics arguments~\cite{tsallis_1988}, an one-parameter generalization of the logarithmic and exponential functions has been proposed~\cite{tsallis_qm}. 
This generalization is not univoquous, other one-parameter logarithmic and exponential functions have been proposed~\cite{kaniadakis_2001,PhysRevE.66.056125}.
A two-parameter generalization has also been proposed~\cite{kaniadakis:046128} to englobe all the previous one-parameter definitions.
Here we will consider the one-parameter generalization obtained from non-extensive thermostatistics, which has provided new insights in the treatment of some problems of complex systems~\cite{europhysicsnews_2005}.  
Arithmetical and differential operators have been constructed to recover standard properties of these functions, such as extensivity~\cite{nivanen_2003,borges_2004,kalogeropoulos_2005}.
Nevertheless, the use of these operators have been restricted to simple cases. 
For instance, on one hand, the consideration of few variables or 
              on the other hand, many compositions of the same variable.  

In this letter we obtain the generalization of the logarithmic and the exponential functions using a simple geometrical argument and introduce the generalization of the arithmetical operators. 
In the following we extend the definition of the sum and product operators to an arbitrary number of terms or factors. 
We show that these generalized sum and product operators are related to the arithmetical and geometrical means of a random variable, which can be of interest in the study of non-equilibrium physical situations.
Using symmetrical polynomials, mixtures of the arithmetical and geometrical means are explicitly studied in an alternative deduction of the successive $\tilde{q}$-sum result. 
The problem of composing interest rates in financial mathematics is an application where the redefinition of the sum operator is interesting since it becomes extensive, as it is shown. 
 
\section{Generalized Functions and Operators}

In the following, we present a brief review of the one-parameter generalization of the logarithmic and exponential functions. 
Also the definitions and properties of the newly defined arithmetical and differential operators are presented. 


The one-parameter generalization of the logarithmic and exponential functions have been proposed~\cite{tsallis_qm}, based on non-extensive thermostatistics arguments~\cite{tsallis_1988}. 
This can be obtained using simple geometrical arguments if one considers the area underneath a non-symmetric hyperbole.


The $\tilde{q}$-logarithmic function $\ln_{\tilde{q}}(x)$ is defined as the value of the area underneath $f_{\tilde{q}}(t)=1/t^{1-\tilde{q}}$ in the interval $t \in [1,x]$:
\begin{eqnarray}
\nonumber
\ln_{\tilde{q}}(x) & = & \int_1^x \frac{\mbox{d}t}{t^{1-\tilde{q}}} = \lim_{\tilde{q}' \rightarrow \tilde{q}}\frac{x^{\tilde{q}'} - 1}{\tilde{q}'} \\ 
                   & = & \left\{
\begin{array}{ll}
\frac{x^{\tilde{q}}-1}{\tilde{q}}, & \mbox{if } \tilde{q} \neq 0, \\
\ln x, & \mbox{if } \tilde{q} = 0.
\end{array}
\right. \; 
\end{eqnarray}
This function is \emph{not} the logarithmic function in the basis $\tilde{q}$ [$\log_{\tilde{q}}(x)$], but a generalization of the natural logarithmic function definition, which is recovered for $\tilde{q} = 0$.  
The area is considered negative for $0 < x < 1$, it vanishes for $x = 1$ and it is positive for $x > 1$, independently of $\tilde{q}$.
Due to the symmetrical aspects when changing $\tilde{q}$ into $-\tilde{q}$, here we have used the representation of Refs.~\cite{nivanen_2003,kalogeropoulos_2005} ($\tilde{q} = a = k$) instead of the traditional nonextensive one ($\tilde{q} = 1 - q$) of Ref.~\cite{borges_2004}.  
This proposed generalization is precisely the form proposed by Montroll and Badger~\cite{badger_1974} to unify the Verhulst and Gompertz one-specie population dynamics as shown in Ref.~\cite[p. 83]{montroll_west}.


In the other way around, given the area $x$ underneath the curve of $f_{\tilde{q}}(t)=1/t^{1-\tilde{q}}$ for $t \in [0,y]$, the upper limit $y$ is given by the generalized $\tilde{q}$-exponential function: $y = e_{\tilde{q}}(x)$.
This is the inverse of the $\tilde{q}$-logarithmic function $e_{\tilde{q}}[\ln_{\tilde{q}}(x)] = x = \ln_{\tilde{q}}[e_{\tilde{q}}(x)]$ and it is given by: 
\begin{eqnarray}
\nonumber
e_{\tilde{q}}(x) & = & 
\left\{
\begin{array}{ll}
\lim_{\tilde{q}' \rightarrow \tilde{q}}[1 + \tilde{q}'x]^{1/\tilde{q}'} & ,  \mbox{if} \;  \tilde{q} x \ge -1 \\
0                                                 & , \mbox{otherwise}
\end{array}
\right.   \\
                 & = & \lim_{\tilde{q}' \rightarrow \tilde{q}}[1 + \tilde{q}'x]_{+}^{1/\tilde{q}'} \; , 
\label{eq:q_exp}
\end{eqnarray}
where the use of the operator $[a]_+ = \max(a,0)$ is necessary since $e_{\tilde{q}}(x)$ is not real if $\tilde{q} x < -1$.  
This is a non-negative function $e_{\tilde{q}}(x) \ge 0$, with $e_{\tilde{q}}(0) = 1$, for any $\tilde{q}$.
From the definition of the $\tilde{q}$-exponential function, other functions can be $\tilde{q}$ generalized such as the hyperbolic functions~\cite{borges_1998}, trigonometric function~\cite{borges_1998}, gaussian probability distribution function~\cite{PhysRevLett.75.3589} etc.


In the following we show that standard composition rules (extensivity properties) can be reobtained generalizing algebraic operators~\cite{nivanen_2003,borges_2004}.  


The definition of the $\tilde{q}$-addition operator of two real numbers is: 
\begin{equation}
a \oplus_{\tilde{q}} b = a + b + \tilde{q} a b \; ,
\end{equation} 
so that one recovers usual summation: $\oplus_0 = +$, for $\tilde{q} = 0$.

In this way, one has the following composition rules:
\begin{eqnarray}
\ln_{\tilde{q}}(x_1 \cdot x_2) & = & \ln_{\tilde{q}}(x_1) \oplus_{\tilde{q}} \ln_{\tilde{q}}(x_2) \; \; \; \; \; \mbox{and} \\
e_{\tilde{q}}(x_1) \cdot e_{\tilde{q}}(x_2) & = & e_{\tilde{q}}(x_1 \oplus_{\tilde{q}} x_2) \; . 
\end{eqnarray}
Associativity [$(a \oplus_{\tilde{q}} b) \oplus_{\tilde{q}} c = a \oplus_{\tilde{q}} (b \oplus_{\tilde{q}} c)$] and 
commutativity ($a \oplus_{\tilde{q}} b = b \oplus_{\tilde{q}} a$) are properties of this operator. 
Nevertheless, some care must be taken with the ditributive property since $c (a \oplus_{\tilde{q}} b) = c a + c b + \tilde{q} c a b \neq (c a) \oplus_{\tilde{q}} (c b)$.


The $\tilde{q}$-multiplication operator is defined as: 
\begin{equation}
a \otimes_{\tilde{q}} b = [a^{\tilde{q}} + b^{\tilde{q}} - 1]_{+}^{1/\tilde{q}} \; , 
\end{equation}
so that one has the usual product: $\otimes_0 = \times$, for $\tilde{q} = 0$.

This product leads to the following composition rules: 
\begin{eqnarray}
\ln_{\tilde{q}}(x_1 \otimes_{\tilde{q}} x_2) & = & \ln_{\tilde{q}}(x_1) + \ln_{\tilde{q}}(x_2) \; \; \; \; \; \mbox{and} \\
e_{\tilde{q}}(x_1) \otimes_{\tilde{q}} e_{\tilde{q}}(x_2) & = & e_{\tilde{q}}(x_1 + x_2) \; .
\end{eqnarray}
Associativity [$(a \otimes_{\tilde{q}} b) \otimes_{\tilde{q}} c = a \otimes_{\tilde{q}} (b \otimes_{\tilde{q}} c)$] and commutativity ($a \otimes_{\tilde{q}} b = b \otimes_{\tilde{q}} a$) are properties of this operator. 
Here again, some care must be taken with ditributive property since $(a \otimes_{\tilde{q}} b)^c = [a^{\tilde{q}} + b^{\tilde{q}} - 1]_{+}^{c/\tilde{q}} \neq a^c \otimes_{\tilde{q}} b^c$. 


These results lead to the following properties:
\begin{eqnarray}
\ln_{\tilde{q}}\left( \prod_{i = 1}^n \otimes_{\tilde{q}} \, x_i  \right) & = & \sum_{i=1}^n \ln_{\tilde{q}} x_i \\
e_{\tilde{q}}\left( \sum_{i = 1}^n \oplus_{\tilde{q}} \, x_i \right) & = & \prod_{i=1}^{n} e_{\tilde{q}}(x_i) \; .
\end{eqnarray}

Although the $\tilde{q}$-subtraction and $\tilde{q}$-division operators will not be addressed in this letter, for completeness we will present them.  


The neutral element for the $\tilde{q}$-adition is 0, so that the inverse element $\ominus_{\tilde{q}} a$ of $a \ne -1/\tilde{q}$ is obtained imposing $a \oplus_{\tilde{q}} (\ominus_{\tilde{q}} a) = 0$, which leads to: $\ominus_{\tilde{q}} a = -a/(1 + \tilde{q} a)$. 
In this way one writes the generalized subtraction operator as: 
\begin{equation}
b \ominus_{\tilde{q}} a = b \oplus_{\tilde{q}} (\ominus_{\tilde{q}} a) = (b-a)/(1 + \tilde{q} a) \; ,
\end{equation}
which leads to :  
\begin{eqnarray}
\ln_{\tilde{q}}( x_1/x_2) & = & \ln_{\tilde{q}}(x_1) \ominus_{\tilde{q}} \ln_{\tilde{q}}(x_2) \; \; \; \; \; \mbox{and}\\  
e_{\tilde{q}}(x_1)/e_{\tilde{q}}(x_2) & = & e_{\tilde{q}}(x_1 \ominus_{\tilde{q}} x_2) \; .
\end{eqnarray}


The neutral element for the $\tilde{q}$-multiplication is 1, so that the inverse element $1 \oslash_{\tilde{q}} a$ of $a \ge 0$ is obtained imposing $a \otimes_{\tilde{q}} (1 \oslash_{\tilde{q}} a) = 1$, which leads to:  $1 \oslash_{\tilde{q}} a = [2 - a^{\tilde{q}}]_+^{1/\tilde{q}}$.
Observe that $1 \oslash_{\tilde{q}} 0 $ does not diverge if $\tilde{q} \ne 0$. 
In this way one defines: 
\begin{equation}
b \oslash_{\tilde{q}} a = b \otimes_{\tilde{q}} (1 \oslash_{\tilde{q}} a) = (b^{\tilde{q}} - a^{\tilde{q}} + 1)^{1/\tilde{q}} \;, 
\end{equation}
where $a, b > 0$. 
With this division operator one retrieves the following extensivity properties: 
\begin{eqnarray}
\ln_{\tilde{q}}(x_1 \oslash_{\tilde{q}} x_2) & = & \ln_{\tilde{q}}(x_1) - \ln_{\tilde{q}}(x_2) \; \; \; \; \; \mbox{and}\\ 
e_{\tilde{q}}(x_1) \oslash_{\tilde{q}} e_{\tilde{q}}(x_2) & = & e_{\tilde{q}}(x_1 - x_2) \;. 
\end{eqnarray}

Moreover, we observe that:
\begin{enumerate}
\item if $0 \le a^{\tilde{q}} \le 2$ then $1 \oslash_{\tilde{q}} (1 \oslash_{\tilde{q}} a) = a$,  
\item if $\tilde{q} > 0$ then $1 \oslash_{\tilde{q}} 0 = 2^{1/\tilde{q}}$, 
\item if $a^{\tilde{q}} \le 1 + b^{\tilde{q}}$ then $a \oslash_{\tilde{q}} b = 1 \oslash_{\tilde{q}} (b \oslash_{\tilde{q}}  a )$ and 
\item if $c^{\tilde{q}} -1 \le b^{\tilde{q}} \le a^{\tilde{q}} + 1$ then $a \oslash_{\tilde{q}} (b \oslash_{\tilde{q}} c) = (a \oslash_{\tilde{q}} b) \otimes_{\tilde{q}} c = (a \otimes_{\tilde{q}} c) \oslash_{\tilde{q}} b$.
\end{enumerate}

\section{Generalized Successive Sum and Product Operators}

So far we have dealt with the composition of only two quantities ($a$ and $b$).
We proceed with the goal to calculate the successive sum and product of those generalized operators to the $n$ values $x_1, x_2, \ldots, x_n$ of the quantity named $X$ and interpret our findings. 

\subsection{Mean Values}

Initially we recall that the arithmetical mean, which is appropriate for additive processes, is defined as: $\langle X \rangle_{A} = (\sum_{i=1}^n x_i)/n$ and the geometrical mean, which is appropriate for multiplicative processes, is:
$\langle X \rangle_{G} = (\prod_{i=1}^n x_i)^{1/n}$.


Let us consider the arithmetical mean  of the $\tilde{q}$-logarithm of $X$,
\begin{equation}
\langle \ln_{\tilde{q}}(X) \rangle_{A} = \frac{1}{n} \sum_{i=1}^n \ln_{\tilde{q}}(x_i) = \frac{1}{n} \ln_{\tilde{q}} \left( \prod_{i = 1}^n \otimes_{\tilde{q}} \, x_i \right) \; ,
\label{eq:a_m_q_log}
\end{equation}
which is equal to the $\tilde{q}$-logarithm of the $\tilde{q}$-product over the number of terms. 

Statistically, now let $X$ be a random variable:
\begin{eqnarray}
\nonumber
\langle \ln_{\tilde{q}}(X) \rangle_{A} & = & \frac{1}{n} \ln_{\tilde{q}} \left( \prod_{i = 1}^n \otimes_{\tilde{q}} \, x_i \right) \\ 
\nonumber
                                       & = & \frac{1}{n} \ln_{\tilde{q}} \left( \left[ \sum_{i=1}^{n} x_i^{\tilde{q}} - (n - 1) \right]^{1/\tilde{q}} \right)  \\ 
                                       & = &\frac{\langle X^{\tilde{q}}\rangle_A - 1}{\tilde{q}}\; .
\end{eqnarray}
In this way we see that the $\tilde{q}$th moment of the quantity $X$ is related to the arithmetical mean of the $\tilde{q}$-logarithm:
\begin{equation}
\langle X^{\tilde{q}}\rangle_A = 1 + \tilde{q} \langle \ln_{\tilde{q}}(X) \rangle_{A} \; .
\end{equation}


The geometrical mean of the $\tilde{q}$-exponential of $X$ is:
\begin{eqnarray}
\nonumber
\langle e_{\tilde{q}}(X) \rangle_{G} & = & \left[ \prod_{i=1}^n e_{\tilde{q}}(x_i) \right]^{1/n} \\
                                     & = & \left[ e_{\tilde{q}}\left( \sum_{i = 1}^n \otimes_{\tilde{q}} x_i \right) \right]^{1/n}\; , 
\label{eq:q_m_q_exp}
\end{eqnarray}
which is related to the $\tilde{q}$-exponential of the $\tilde{q}$-sum raised to the inverse of the number of factors.

Notice that for $\tilde{q}=0$ the following relationship between these two means holds:
$\ln \langle x \rangle_{G} =  \langle \ln x \rangle_{A}$ and $\langle e^x \rangle_{G} =  e^{\langle x \rangle_{A}}$.

\subsection{Successive Operations}

To obtain the successive generalized addition operator, we use the result of Eq.~\ref{eq:a_m_q_log} to define the $\tilde{q}$-sum operator:


\begin{eqnarray}
\nonumber
\sum_{i = 1}^n \oplus_{\tilde{q}} \, x_i & = & x_1 \oplus_{\tilde{q}} x_2 \oplus_{\tilde{q}} \cdots \oplus_{\tilde{q}} x_n 
                                            = \ln_{\tilde{q}} \left[ \langle e_{\tilde{q}}(X) \rangle_{G}^{n} \right]  \\
                                          & = & \frac{1}{\tilde{q}} \left[ \prod_{i=1}^{n} (1 + \tilde{q} x_i) - 1 \right] \; .
\label{eq:qsoma}
\end{eqnarray}


To define the successive generalized multiplication operator, i.e., $\tilde{q}$-product operator, we use the result of Eq.~\ref{eq:q_m_q_exp} is:
\begin{eqnarray}
\nonumber
 \prod_{i = 1}^n \otimes_{\tilde{q}} \, x_i & = & x_1 \otimes_{\tilde{q}} x_2 \otimes_{\tilde{q}} \cdots \otimes_{\tilde{q}} x_n 
                                              =  e_{\tilde{q}} \left[ n \langle \ln_{\tilde{q}}(X) \rangle_{A} \right] \\ 
                                             & = & \left[ \sum_{i=1}^{n} x_i^{\tilde{q}} - (n - 1) \right]^{1/\tilde{q}} \; .
\end{eqnarray}


The validity of these results can be checked by induction and particular cases can be obtained. 

It is not difficult to show that for $\tilde{q} = 0$, one retrieves standard results: 
\begin{equation}
\sum_{i = 1}^n \oplus_{0} \, x_i  =  \sum_{i = 1}^n x_i 
\; \; \; \; \;  \mbox{and} \; \; \; \; \; 
\prod_{i = 1}^n \otimes_{0} \, x_i  =  \prod_{i = 1}^n  x_i \, . 
\end{equation}

The $n$ times $\tilde{q}$-addition of the same term leads to:
\begin{eqnarray}
 \sum_{i = 1}^n \oplus_{\tilde{q}} \, a & = & \frac{[1 + \tilde{q} a]^n - 1}{\tilde{q}} = \ln_{\tilde{q}} \{ [ e_{\tilde{q}}(a) ]^ n \} \; ,  
\label{eq_n_q_sum}
\end{eqnarray}
which has been previously found in Ref.~\cite{borges_2004}.

The $\tilde{q}$-product of $n$ times the same factor leads to $\tilde{q}$-power operator $\wedge_{\tilde{q}}$:
\begin{eqnarray}
\nonumber
a \wedge_{\tilde{q}} n & = &   \prod_{i = 1}^n \otimes_{\tilde{q}} \, a = [n a^{\tilde{q}} - (n-1)]^{1/\tilde{q}}  \\ 
                       & = & e_{\tilde{q}}[ n \ln_{\tilde{q}}(a) ] \; .  
\end{eqnarray}
This operator and result have also been previously found in Ref.~\cite{borges_2004}. 
It is interesting to point out the  following fundamental limit~\cite{suyari_2006}:
\begin{equation}
\lim_{n \rightarrow \infty} \left(1 + \frac{a}{n} \right) \wedge_{\tilde{q}} n = e_{\tilde{q}}(a) \; .
\end{equation}

\subsection{Symmetrical Polynomials}

There is an alternative manner to calculate the $\tilde{q}$-sum operator. 
This different derivation stresses some characteristics of this operator.
If one expands the $\tilde{q}$-sum operator, one gets a polynomial of order $n-1$ in $\tilde{q}$, where the coefficient of $\tilde{q}^k$ is  related to $k + 1$ moment:
\begin{eqnarray}
\sum_{i = 1}^n \oplus_{\tilde{q}} \, x_i & = &  \sum_{i_1 = 1}^n x_{i_1} + 
\tilde{q} \sum_{i_1 = 1}^n x_{i_1}  \sum_{i_2 = i_1 + 1}^n  x_{i_2} + \nonumber \\ 
&   & \tilde{q}^2 \prod_{j = 1}^3 \sum_{i_j = i_{j-1} + 1 }^n x_{i_j}  + 
      \cdots + \tilde{q}^{n-1} \prod_{i=1}^n x_i \nonumber \\ 
& = & \sum_{k=0}^{n-1} \tilde{q}^k \prod_{j = 1}^{k+1} \sum_{i_j = i_{j-1} + 1 }^n x_{i_j} \; ,
\label{eq:aux1}
\end{eqnarray} 
where $i_0 = 0$ and we clearly see that the coefficient of $\tilde{q}^0$ is proportional to the arithmetical mean and the one of $\tilde{q}^{n-1}$ is related to the geometrical mean of $X$. 
Intermediary powers of $\tilde{q}$ translates as a mixture of the arithmetical and geometrical means, where the proportion of each one is given by the power of $\tilde{q}$.  
This mixture of arithmetical with geometrical means also appears when dealing with polynomials. 
Consider a generic polynomial of degree $n$ in $\lambda$ written in its factored form (as a function of its $n$ roots $x_i$): $p(\lambda)  =  \prod_{i=1}^n ( \lambda - x_i)$ and its development: 
\begin{eqnarray*}
p(\lambda) & = & \lambda^n - \lambda^{n-1} \sum_{i_1 = 1}^n x_{i_1} + 
                 \lambda^{n-2} \sum_{i_1 < i_2} x_{i_1} x_{i_2} - \\  
           &   & \lambda^{n-3} \sum_{i_1 < i_2 < i_3} x_{i_1} x_{i_2} x_{i_3} + \cdots + \\
           &   & (-1)^n \prod_{i_n = 1}^n x_{i_n} \; .
\end{eqnarray*} 
The coefficients of $\lambda$ are the ones found in the $\tilde{q}$-sum. 
A perfect equivalence is obtained for  $\lambda \ne 0$, which leads to:
\begin{eqnarray}
\nonumber
\lambda - \frac{p(\lambda)}{\lambda^{n-1}} & = &\sum_{i_1 = 1}^n x_{i_1} - \lambda^{-1} \sum_{i_1 < i_2} x_{i_1} x_{i_2} + \\
                                           &   & \lambda^{-2} \sum_{i_1 < i_2 < i_3} x_{i_1} x_{i_2} x_{i_3} + \cdots \nonumber \\ & & 
+ (-1)^{n-1} \lambda^{1-n} \prod_{i_n = 1}^n x_{i_n} \; .
\label{eq:aux2}
\end{eqnarray}
Comparing Eqs.~\ref{eq:aux1} and~\ref{eq:aux2}, one obtains $\lambda^{-1} = - \tilde{q}$.
So, all coefficients in Eq.~\ref{eq:aux2} become positive and one rewrites Eq.~\ref{eq:aux1} as:
\begin{eqnarray*}
\sum_{i = 1}^n \oplus_{\tilde{q}} \, x_i & = &  -\frac{1}{\tilde{q}} - (-\tilde{q})^{n-1} p(-1/\tilde{q}) \; , 
\end{eqnarray*}
which leads to the result of Eq.~(\ref{eq:qsoma}).

\section{Applications}

In the following we consider two applications. 
In the first one we show a simple way to obtain Tsallis entropy.
In the second example we extend the procedure pointed out in Ref.~\cite{nivanen_2003} when composing interest rates into a saving  account to an arbritary number of time intervals. 
Also we consider the situation of aporting or withdrawing money from that saving account.    

\subsection{Generalized Entropy}

Now, let us compute the mean value of $\ln_{\tilde{q}} (1/p_i)$, for a discrete system with $W$  available states and $p_i$ is the probability of accessing state $i$~\cite{tsallis_2004}. 
The arithmetical mean of the $\tilde{q}$-logarithm of the inverse probability (surprise) is postulated to be proportional to the Tsallis entropy $S_{\tilde{q}}$~\cite{tsallis_1988,tsallis2}. 
\begin{eqnarray}
\nonumber
\langle \ln_{\tilde{q}} \frac{1}{p_i} \rangle_A & = & \sum_{i=1}^W p_i \ln_{\tilde{q}} \frac{1}{p_i} = \sum_{i=1}^W p_i \frac{p_i^{-\tilde{q}} - 1}{\tilde{q}} \\ 
                                              & = & \frac{\sum_{i=1}^W p_i^{1 - \tilde{q}}}{\tilde{q}} = \frac{S_{\tilde{q}}}{K_B} \; ,
\end{eqnarray} 
where $K_B$ is the Boltzmann's constant. 
The $S_{\tilde{q}}$ entropy reduces to Boltzmann-Gibbs entropy for $\tilde{q} = 0$ and if 
all states $i$ are equally accessible, then $p_i = 1/W$ and $S_{\tilde{q}}/K_B = \ln_{\tilde{q}} W$. 

\subsection{Composing Interest Rates} 

Consider now the evolution of capital in a savings account with an initial amount $x_0$ subjected to interest rates $r_i$ at time intervals $\Delta t$.  
After the first time interval $\Delta t$, the amount is $x_1 = x_0 (1 + r_1)$ and 
after two time intervals, the amount is $x_2 = x_1 (1 + r_2) = x_0 (1 + r_1)(1 + r_1)$. 
The effective interest rate in this two time intervals is: $r^{(2)} = x_2 / x_0 - 1 = r_1 + r_2 + r_1 r_2 =  r_1 \oplus_1 r_2$.  
Notice that $r^{(1)} = r_1$.

For the $n$th time steps, the initial amount $x_0$ becomes $x_n = x_0 \prod_{i=1}^n (1 + r_i)$. 
The effective interest after $n$ time intervals is:
\begin{equation}
r^{(n)} = \frac{x_n}{x_0} - 1 = \prod(1 + r_i) - 1 = \sum_{i=1}^{n} \oplus_1 r_i \; ,
\end{equation} 
where we have used Eq.~(\ref{eq:qsoma}).
In this way we generalize the result firstly obtained in Ref.~\cite{nivanen_2003}.
Interest rate composition is extensive using the $\tilde{q}$-sum operator for a special value $\tilde{q} = 1$. 


Let the variable $a_i$ represent amount of deposit ($a_i > 0$) or withdraw ($a_i < 0$)  in time interval $i$ in the savings account, so that $a_0 = x_0$.
In this case one has the amount obtained at the $n$-th time step is: 
\begin{equation}
x_n = \sum_{k=0}^{n-1} a_k \; \sum_{i = k+1}^n\oplus_1 r_i \; .
\end{equation} 
In this result one sees the simple ($\tilde{q} = 0$) extensivity of amounts $a_i$ weighted by the interest rates, which are non-trivially extensive ($\tilde{q} = 1$).

\section{Conclusion}

The analytical expressions for the sucessive generalized sum and product operators fulfills an important gap found in the $\tilde{q}$-algebra and may be useful, for instance, when considering averages of annealed out-of-equilibrium systems (averages of generalized Helmholtz potentials). 
This shall be explored in future works. 
Also, as it is well known, interest composition rate is not an additive process. 
Nevertheless if one considers the $\tilde{q}$-sum operator, interest rate composition becomes an additive process for $\tilde{q} = 1$.  

\section*{Acknowledgments}

Authors thank N. A. Alves and A. L. Esp\'indola for fruitful discussions.
ASM acknowledges the Brazilian agencies CNPq (305527/2004-5) and FAPESP (2005/02408-0) for support.  
RSG and TJA also acknowledge CNPq (140420/2007-0) and FAPESP (06/03435-4) respectively for support.


\end{document}